# AI-Based Autonomous Line Flow Control via Topology Adjustment for Maximizing Time-Series ATCs

Tu Lan, Jiajun Duan, Bei Zhang, Di Shi, Zhiwei Wang, Ruisheng Diao, Xiaohu Zhang
GEIRI North America, AI & System Analytics
San Jose, CA, USA
tu.lan@geirina.net; di.shi@geirina.net

*Abstract*—This paper presents a novel AI-based approach for maximizing time-series available transfer capabilities (ATCs) via autonomous topology control considering various practical constraints and uncertainties. Several AI techniques including supervised learning and deep reinforcement learning (DRL) are adopted and improved to train effective AI agents for achieving the desired performance. First, imitation learning (IL) is used to provide a good initial policy for the AI agent. Then, the agent is trained by DRL algorithms with a novel guided exploration technique, which significantly improves the training efficiency. Finally, an Early Warning (EW) mechanism is designed to help the agent find good topology control strategies for long testing periods, which helps the agent to determine action timing using power system domain knowledge; thus, effectively increases the system error-tolerance and robustness. Effectiveness of the proposed approach is demonstrated in the "2019 Learn to Run a Power Network (L2RPN)" Global Competition, where the developed AI agents can continuously and safely control a power grid to maximize ATCs without operator's intervention for up to 1-month's operation data and eventually won the first place in both development and final phases of the competition. The winning agent has been open-sourced on GitHub.

## I. Introduction

Maximizing available transfer capabilities (ATCs) is of critical importance to bulk power systems from both security and economic perspectives, which represents the remaining transfer margin of transmission network for further energy transactions. Due to environmental and economic concerns, transmission expansion via building new lines for enlarging transfer capabilities is no longer an easy option for many utilities across the world. Additionally, the increasing penetration of renewable energy, demand response, electric vehicles, and power-electronics equipment has caused more stochastic and dynamic behavior that threatens safe operation of the modern power grid [1]-[2]. Thus, it becomes essential to develop fast and effective control strategies for maximizing ATCs considering uncertainties while satisfying various security constraints.

Compared with re-dispatching generators, shedding electricity demands, and installing FACTS devices, active network topology control via transmission line switching or bus splitting for increasing ATCs and mitigating congestions provides a low-cost and effective solution, especially for a deregulated power market or utilities with limited choices (e.g., RTE France with nuclear power supplying vast majority of its demands). This idea was first proposed in the early 1980s when several research efforts were conducted for achieving multiple control purposes such as cost minimization, voltage, and line flow regulation [3]-[4]. Transmission line switching or bus splitting/rejoining is essentially a multivariate discrete programming problem that is difficult to solve, given the complexity and uncertainties of bulk power systems. Various approaches have been reported to tackle this problem. In [5], a mixed-integer linear programming (MIP) model is proposed with DC power flow approximation of the power network, where the generalized optimization solver, CPLEX, is adopted to solve the MIP. In [6], the transmission switching (TS) optimization process with DCOPF is decoupled from a master unit commitment procedure, where the optimal TS schedule is formulated as a MIP problem that is again solved using CPLEX. Reference [7] presents a fast heuristic method to speed up the convergence using the aforementioned modeling and solution practice. Similar approaches with variations are also reported in [8] and [9], which use a point estimation method for modeling system uncertainties with AC power flow feasibility checking and correction modules.

However, several limitations are observed in existing methods, including: (a) Linear approximation in DC power flow without considering all security constraints is typically utilized, which affects the solution accuracy for a real-world power grid. Using full AC power flow with all security constraints for optimization becomes non-convex due to the high nonlinear nature of power grids, which cannot be effectively solved using state-of-the-art techniques without relaxing/sacrificing certain security constraints or solution accuracy. (b) The combination set of lines and bus-bars to be switched simultaneously grows exponentially; in addition, sensitivity-based methods are susceptible to changing system operating conditions. Thus, it may take a long time to solve such an optimization process for a large power grid, preventing the solution from being deployed in the real-time environment.

To fill these technology gaps, this research presents a novel method that adopts AI-based algorithms (IL and DRL) with several innovative techniques (including guided exploration and early warning) for training effective agents in providing fast and autonomous topology control strategies for maximizing time-series ATCs. The developed techniques were used to participate in the *2019 L2RPN*, a global power system AI competition hosted by RTE France and ChaLearn

---

This work was supported by SGCC Science and Technology Program.

[10], considering full AC power flow and practical constraints, which eventually outperformed all competitors' algorithms. The remainder of the paper is organized as follows: section II presents the problem formulation and introduces the principle of reinforcement learning for solving the Markov Decision Process (MDP). Section III provides the detailed architecture design, key steps, AI algorithms with several innovative techniques, and implementation of the proposed methodology for autonomous topology control. Case studies are presented in section IV to demonstrate the effectiveness of the proposed method. Finally, conclusions are drawn in section V with future work discussed.

## II. PROBLEM FORMULATION

### A. Objectives, control measures, and practical constraints

The problem to solve in this research is discussed in the *2019 L2RPN* challenge with full details [10]. The main objective is to maximize the ATCs of a given power grid over all time steps of various scenarios. Each scenario is defined as operating the grid for a consecutive time period, e.g., four weeks with a fixed time interval of 5 minutes, considering daily load variations, pre-determined generation schedules and real-time adjustment, voltage setpoints of generator terminal buses, network maintenance schedules and contingencies. The control decisions only include network topology adjustment, namely, one node splitting/rejoining operation, one line switching, and the combination of these two. System generation and loads are not allowed to be controlled for enhancing ATCs. Several hard constraints are considered for all the scenarios of interest: (a) system demands should be met at any time without load shedding; (b) no more than one power plant can be tripped; (c) no electrical islands can be formed as a result of topology control; (d) AC power flow should converge at all time. It will cause "game over" if any hard constraint is violated. For soft constraints, violations lead to certain consequences instead of immediate "game over". Overloaded lines over 150% of their ratings are tripped immediately, which can be recovered after 50 minutes (10 time steps); while for overloaded lines below 150% of their ratings, control measures can be used to mitigate the overloading issue with a time limit of 10 minutes (2 time steps). If still overloaded, the line will be tripped, and cannot be recovered until after 50 minutes. In addition, a practical constraint is considered that is to allow a "cooldown time" (15 minutes) before a switched line or node can be reused for action. Both soft and hard constraints make the problem more practical and close to real-world grid operation. To examine the performance of agents, metrics in Eq. (1) are used, which measure the time-series ATCs for a power grid.

$$\begin{aligned} step\_score &= \sum_{i=1}^{n\_lines} max(0, \ 1 - (\tfrac{lineflow_i}{thermallimit_i})^2) \\ chronic\_score &= \begin{cases} 0 & if \ gameover \\ \sum_{j=1}^{n\_steps} step\_score_j & otherwise \end{cases} \\ total\_score &= \sum_{k=1}^{n\_chronics} chronic\_score_k \end{aligned} \quad (1)$$

The detailed mathematical formulation can be found in [11] and therefore is not repeated here due to space limitation.

### B. Problem formulated as MDP

Maximizing time-series ATCs via topology control or adjustment can be modeled as an MDP [12], which consists of 5 key elements: a state space $\mathcal{S}$, an action space $\mathcal{A}$, a transition matrix $\mathcal{P}$, a reward function $\mathcal{R}$, and a discount factor $\gamma$. In this work, an AC power flow simulator is used to represent the environment [13]. The agent state ($s_t^a \in \mathcal{S}$) is a partial observation from the environment state ($s_t^e \in \mathcal{S}$). State $s_t^a$ contains 538 features, including active power outputs and voltage setpoints of generators, loads, line status, line flows, thermal limits, timestamps, etc. The action space $\mathcal{A}$ is formed by including line switching, node splitting/rejoining, and a combination set of both. An immediate reward $r_t$ at each time step is defined in Eq. (2) to assess the remaining available transfer capabilities:

$$r_t = \begin{cases} -1 & if \ game \ over \\ \tfrac{1}{N}\sum_{i=1}^{N} max(0, \ 1 - (\tfrac{lineflow_i}{thermallimit_i})^2) & otherwise \end{cases} \quad (2)$$

In MDP, a cumulative future return $R_t$ is defined which contains the immediate reward and the discounted future rewards, defined in Eq. (3) [12]:

$$R_t = r_t + \gamma r_{t+1} + \ldots + \gamma^T r_{t+T} = \sum_{k=0}^{T} \gamma^k r_{t+k} \quad (3)$$

where $T$ is the length of the MDP chain, and $\gamma \in [0, 1]$ is the discount factor.

### C. Solving MDP via reinforcement learning

With recent success in various control problems with high nonlinearity and stochastics, reinforcement learning is adopted which exhibits great potentials in maximizing long-term rewards for achieving a specific goal [1]-[2]. Various RL algorithms exist with pros and cons. One typical example is Q-learning, which utilizes a Q-table to map each state and action pair using an action-value, $Q(s,a)$, which evaluates action $a$ taken at state $s$ by considering the future cumulative return $R_t$. According to the Bellman Equation [12], the cumulative return can be represented as an expected return, shown in Eq. (4):

$$\begin{aligned} Q(s,a) &= \mathbb{E}[R_t \mid S_t = s, A_t = a] \\ &= \mathbb{E}[r_t + \gamma Q(S_{t+1}, A_{t+1}) \mid S_t = s, A_t = a] \end{aligned} \quad (4)$$

To obtain the optimal action-value $Q^*(s,a)$, Q-learning looks one step ahead after taking action $a$ at state $s_t$, and greedily considers the action $a_{t+1}$ at state $s_{t+1}$ for maximizing the expected target value $r_t + \gamma Q^*(s_{t+1}, a_{t+1})$. Using the Bellman equation, the algorithm can perform online updates to control the Q-value towards the Q-target.

$$\begin{aligned} Q(s_t, a_t) \leftarrow &\ Q(s_t, a_t) + \\ &\ \alpha[r_t + \gamma \max_{a_{t+1} \in \mathcal{A}} Q(s_{t+1}, a_{t+1}) - Q(s_t, a_t)] \end{aligned} \quad (5)$$

where $\alpha$ represents the learning rate. Using a Q-table, both the state and action need to be discrete, thus making it difficult to handle complex problems. To overcome this issue, the deep Q network (DQN) method was developed which uses neural networks as a function approximator to estimate the Q-values, $Q(s,a)$, so it can support continuous states in the RL process without discretization of states or building the Q-table. Weights $\theta$ of the neural network represent the mapping from states to Q-values, and therefore, a loss function $L_i(\theta)$ is needed to update the weights and their corresponding Q-values, using Eq. (6) [14]:

$$L_i(\theta_i) = \mathbb{E}_{s,a \sim \rho(\cdot)}\left[\left(y_i - Q(s,a;\theta_i)\right)^2\right] \quad (6)$$

where $y_i = \mathbb{E}_{s' \sim \varepsilon}[r + \gamma \max_{a'} Q(s', a'; \theta_{i-1}) \mid s, a]$, and $\rho$ is the probability distribution of the state and action pair $(s, a)$. By differentiating the loss function using Eq. (7) and performing stochastic gradient descent, weights of the agent can be updated [14].

$$\nabla_{\theta_i} L_i(\theta_i) = \mathbb{E}_{s,a\sim\rho(\cdot); s'\sim\varepsilon}[(r + \gamma \max_{a'} Q(s',a';\theta_{i-1}) - Q(s,a;\theta_i))\nabla_{\theta_i} Q(s,a;\theta_i)] \quad (7)$$

Given its advantages, DQN is selected as the fundamental DRL algorithm in this work to train AI agents for providing topology control actions. However, overestimation is a well-known and long-standing problem for all Q-learning based algorithms. To address this issue, Double DQN (DDQN) that decouples the action selection and action evaluation using two separate neural networks is proposed in [15]. It demonstrates good performance in overcoming the overestimation problem and can obtain better results on ATARI 2600 games than other Q-learning based methods. In addition, a new model architecture, Dueling DQN is proposed in [16], which decouples a single-stream DDQN into a state-value stream and an action-advantage stream, and therefore, the Q-value can be represented as Eq. (8) [16].

$$Q(s,a;\theta,\alpha,\beta) = V(s;\theta,\beta) + \left(A(s,a;\theta,\alpha) - \frac{1}{|\mathcal{A}|}\sum_{a'} A(s,a';\theta,\alpha)\right) \quad (8)$$

The stand-alone state value stream is updated at each step of training process. The frequently updated state-values and the biased advantage values allow better approximation of the Q-values, which is the key in value-based methods. It allows a more accurate and stable update for the agent. Thus, dueling DQN is selected as the baseline model in this work to achieve good control performance.

## III. THE PROPOSED METHODOLOGIES

### A. Architecture design

The architecture of training DRL agents for maximizing ATCs is shown in Fig. 1, where several novel methods are developed. First, imitation learning is used to generate a good initial policy for the dueling DQN agent so that exploration and training time can be greatly reduced; additionally, the agent is less likely to fall into a local optimum. Second, a guided exploration method is used to train the agent instead of the traditional Epsilon-greedy exploration. Third, importance sampling is used to increase the mini-batch update efficiency [17]. Moreover, an Early Warning (EW) system is designed to increase the system robustness. Details regarding these techniques are discussed in the following subsections.

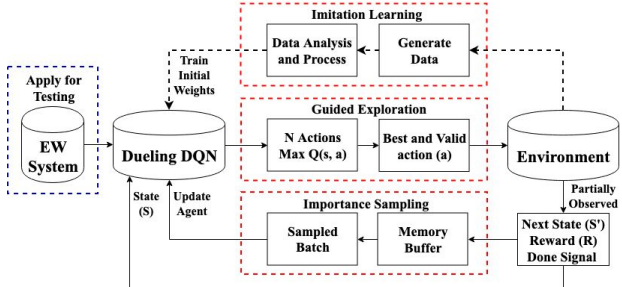

Fig. 1. Overview of the system architecture.

### B. Dueling DQN Agent

The architecture of dueling DQN is given in Fig. 2. The original structure is adopted with a batch normalization layer added to the input layer, and the number of neurons in the hidden layer is modified according to the dimensions of inputs and outputs. The dueling structure decouples the single stream into a state value stream and an advantage stream. The dueling DQN also uses three important techniques in DQN, including: (1) an experience replay buffer that allows the agent to be trained off-policy and decouples the strong correlations between the consecutive training data; (2) importance sampling is used to increase the algorithm learning efficiency and final policy quality [17], by measuring importance of the data using absolute TD-error and giving important data higher priority to be sampled from memory buffer during the training process; and (3) adoption of a DDQN structure, which fixes the q-targets periodically, and then stabilizes the agent updates. The algorithm for training dueling DQN agents is given in Algorithm I.

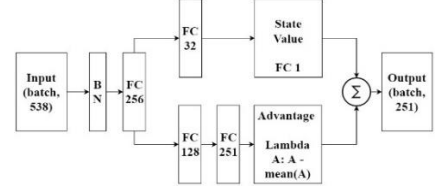

Fig. 2. Architecture of the Dueling DQN.

**Algorithm 1** Double Dueling DQN Guided Exploration Training Method

1: Load pre-trained DQN weights: $\theta = \theta_{init}$.
2: Initialize Memory Buffer D to capacity $N_d$.
3: **for** $episode \leftarrow 1, M$ **do**
4:     Reset the environment, and obtain the state $s_0$
5:     **for** $t \leftarrow 0, T$ **do**
6:        Obtain $Q(\cdot|s_t;\theta)$ from the agent and obtain the set of actions with $N_g$ largest Q values.
7:        Validate and simulate the actions in the set and choose the valid action $a_t$ with best reward.
8:        Execute action $a_t$ in the environment and observe next state $s_{t+1}$, reward $r_t$, and $d_t$.
9:        Store the experience $(s_t, a_t, r_t, s_{t+1}, d_t)$ in D, if $d_t$ is True, store multiple times.
10:       Sample a minibatch of $N_b$ experience $(s_t, a_t, r_t, s_{t+1}, d_t)$ from D using importance sampling.
11:       Calculate q-targets:

$$y_i = \begin{cases} r_i & \text{if } d_t \text{ is True,} \\ r_i + \gamma \max_{a'} Q(s_{t+1}, a'; \theta^-) & \text{otherwise} \end{cases}$$

12:       Update main network using loss function every $N_s$ step: $L_i(\theta) = (y_i - Q(s_t, a_t;\theta))^2$
13:       Hard copy main network weights $\theta$ to the target network $\theta^-$.
14:       Set state $s_t = s_{t+1}$.
15:     **end for**
16: **end for**

### C. Imitation Learning

Imitation learning is essentially a supervised learning method that is used to pre-train DRL agents by providing good initial policies in the form of neural network weights. A power grid simulator is used to generate massive data sets, which are then further processed before being used to train the DQN agent. This process allows the RL agent to obtain good $Q(s, a)$ distributions regarding different input states. The loss function used to train the agent is defined as weighted Mean-Squared-Error (MSE), in Eq. (9):

$$J_\theta = \alpha \times \frac{1}{N}\sum_{i=1}^{N}(Q(s,a_i) - \hat{Q}(s,a_i))^2 + \beta \times \frac{1}{|\mathcal{A}|-N}\sum_{i=N+1}^{|\mathcal{A}|}(Q(s,a_i) - \hat{Q}(s,a_i))^2 \quad (9)$$

where $\alpha, \beta \in [0,1]$, $\alpha + \beta = 1$, $|\mathcal{A}|$ is the size of action space, and vector $\mathbf{Q}(s,a) = [Q(s,a_i), i = 1, ..., |\mathcal{A}|]$ is sorted in descending order. The loss function $J_\theta$ gives a higher weight to actions resulting in high scores, which makes the agent more sensitive to score peaks during the training process, and therefore helps the agent better extract good actions.

*D. Guided exploration training method*

Imitation learning provides a good initial policy for snapshots, and then DRL is used to train the agent for long-term planning capability and to obtain a globally-concerned policy. For DRL training in this problem, the traditional Epsilon-greedy exploration method is inefficient. First, the action space is pretty large and the MDP chain is long. Second, the agent is easy to fall into a local optimum. Thus, a guided exploration method is developed, where actions with the $N_g$ highest Q-values are selected at every timestep, the performance of which are simulated and evaluated on the fly. Then, the action with the highest reward is chosen for implementation and such experience will be stored in the memory. The guided exploration helps the agent to further extract out the good actions. With the help of the action simulation function, the training process is more stable, and better experience is stored and used to update the agent. Thus, guided-exploration significantly increases the training efficiency.

*E. Early warning*

Power systems are highly sensitive to various operating conditions, especially with major topology changes. One bad action may have a long-term adverse effect since the system topology control is successive in a long period of time. The trained DRL agent is not guaranteed to provide a good action every time at various complex system states. Thus, an adaptive mechanism, named Early Warning, is developed in this work which can help the agent determine when to apply action and simulate more actions with high $Q(s,a)$ values to increase the error-tolerance and enhance system robustness, with Fig. 3 illustrating its operation logic.

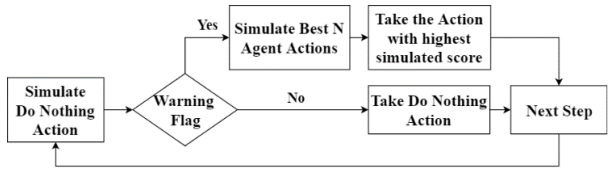

Fig. 3. Early Warning (EW) system workflow.

Initially, at every timestep, it simulates the result of taking no action to the environment, using a warning flag (WF) defined in Eq. (10).

$$WF = \begin{cases} True & \text{if } \frac{lineflow_i}{thermallimit_i} > \lambda, \ \forall i \in \{1,2,\dots,20\} \\ False & \text{otherwise} \end{cases} \quad (10)$$

If the loading level of a line is higher than a pre-determined threshold $\lambda$, a WF is raised. As a result, the $N_g$ top-scored actions are provided by the agent for further simulation. Consequently, the best action with the highest score will be taken. Both guided exploration and the early warning mechanism improve the performance and robustness of the proposed RL algorithm.

## IV. CASE STUDIES AND DISCUSSION

*A. Environment and framework*

A power grid simulator, Python Power Network (Pypownet) [13], is adopted to represent the environment for training RL agents, which is built upon the MATPOWER open-source tool for power grid simulations. It is able to emulate a large-scale power grid with various operating conditions that supports both AC and DC power flow solutions. The framework is developed in Linux, with an interface designed and provided for Reinforcement Learning. The RL agents are trained and tuned using python scripts through massive interactions with Pypowernet. Besides, a visualization module is provided for the users to visualize the system operating status and evaluate control actions in real-time. Several power system models have been provided in this framework with datasets representing realistic time-series operating conditions. The dataset for the IEEE 14-bus model contains 1,000 scenarios with data for 28 continuous days. Each scenario has 8,065 time steps, each representing a 5-minute interval. All models and associated datasets can be directly downloaded from [10].

With the developed environment and framework, the IEEE 14-bus system with the supporting dataset is used to test performance of the proposed DRL agents in autonomous network topology control over long time-series scenarios. In this system, there are a total of 156 different node splitting actions and 20 line switching actions. Thus, an action space of 3,120 is formed by considering null action and all combinations of one node splitting and one line switching without those that can create islands. The DRL agents are trained using Python 3.6 scripts on a Linux server with 48 CPU cores and 128 GB of memory.

*B. Effectiveness of imitation learning for generating good initial policies*

In the first test, a brute-force method is used to train the agent using randomly initialized neural network weights and the full action space with a dimension of 3,120. As expected, due to the large action space and the long time-sequences, the proposed dueling DQN method didn't work well. To solve this problem, the following process is employed to effectively reduce the action space, which includes: (1) 155 node splitting/rejoining actions, (2) 19 line switching actions, and (3) 76 most effective actions with one bus action and one line switching action, and one do-nothing action. In this way, the action space $\mathcal{A}$ is reduced to 251. Then, the imitation learning method introduced in Section III. C is used to obtain good initial policies. Forty scenarios, each with 1,000 timesteps (instead of 8,065), are used for imitation learning, yielding a total number of 40,000 sample pairs, (state, $Q(s,a)$), which are then separated into a training set (90%) and a validation set (10%). Fig. 4 shows a sample prediction and label using IL. After training 100 epochs with a batch size of 1, the weighted MSE decreased to around 0.05, indicating neural networks can generally catch the peaks and trends, and provide relatively effective actions.

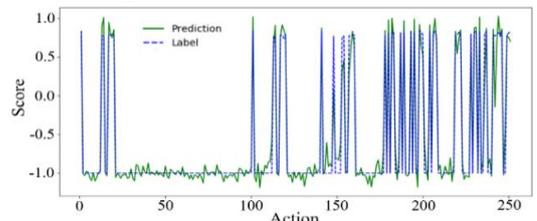

Fig. 4. Sample DQN prediction after imitation learning (loss function: weighted MSE, optimizer: Adam, learning rate: 1e-3).

*C. Improved training performance with guided exploration*

To shorten the MDP chain and decrease the training difficulty, the 28-day scenarios are divided into single days, each with 288 timesteps. For comparison, the training process of dueling DQN agents with Epsilon-greedy exploration and the proposed guided exploration are depicted in Fig. 5(a) and Fig. 5(b), respectively.

With Epsilon-greedy exploration, the agent can hardly control the entire 288 timesteps continuously before Episode 7,000, without game over, although the agent's performance keeps improving towards higher reward values (defined in Eq. (2)). The proposed training process using guided exploration with $N_g$=10 is shown in Fig. 5(b). The agent can control more steps successfully in the earlier phases of the training process compared to Epsilon-greedy exploration. More importantly, it takes a much shorter time to train an agent with a better policy.

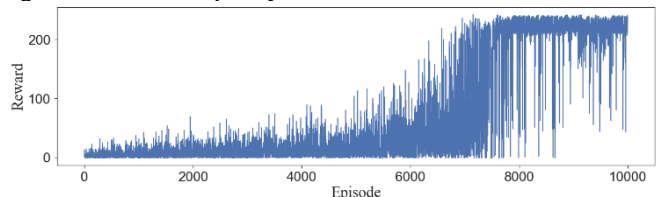

Fig. 5(a). Dueling DQN agent training process using epsilon-greedy exploration.

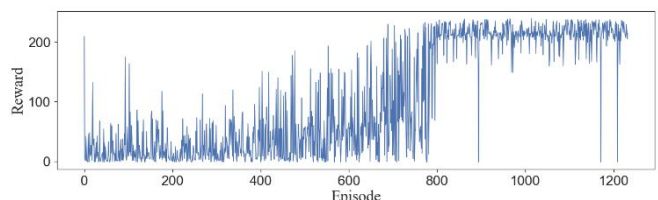

Fig. 5(b). Dueling DQN agent training process using guided exploration.

### D. Testing and performance comparison of different agents

With the proposed methodology, several case studies are conducted with their performance compared in Table I. It is observed that the agent trained only with IL failed for most scenarios. With guided exploration, the agent's performance is greatly improved, where only 7 out of 200 scenarios failed. Using EW (with threshold $\lambda$ ranging from 0.85 to 0.975), the agent can almost handle all the scenarios well with very few cases failed; and the scores are much improved. Similarly, 200 long scenarios with 5,184 time steps are tested using DRL agents, where the best score achieved is 82,687.17, using an EW threshold of 0.93. Only 12 scenarios out of 200 experienced bad control performance. Finally, a well-trained agent was submitted to the *L2RPN* competition with EW $\lambda$=0.885, which was automatically tested using 10 unseen scenarios by the host of the competition, outperformed the other participants, and eventually won the competition. The average decision time for each time step using the proposed agent is roughly 50 ms. The corresponding code and DRL models are open-sourced, which can be found in [18].

## V. CONCLUSION

This paper presents a novel AI-based method to maximize time-series ATCs considering various practical constraints. Several innovative techniques are developed including dueling DQN, imitation learning for generating good initial policies, reduction of action space via simulation and domain knowledge, guided exploration and EW for improving DRL agent's stability and robustness. Massive experiments demonstrate a well-trained AI agent can learn and master the optimal topology control problem for a power grid considering various uncertainties and practical constraints.

Future work will focus on further improving the performance of RL agents, which will be tested on larger power system models. The developed methodologies will also be merged into an AI-based platform developed by the team, Grid Mind [1]-[2], for autonomous grid operation and control.

TABLE I. PERFORMANCE COMPARISON OF DIFFERENT AGENTS ON 200 UNSEEN SCENARIOS WITH 288 TIME STEPS

| Agent | Game Over | Mean Score All | Mean Score w/o Dead |
|---|---|---|---|
| Do Nothing | 91 | 2471.42 | **4534.72** |
| Only Imitation | 198 | 38.21 | 3820.63 |
| Guided Trained | 7 | 4269.63 | 4424.49 |
| EW $\lambda$ = 0.85 | **0** | 4253.40 | 4253.40 |
| EW $\lambda$ = 0.875 | 1 | 4347.56 | 4369.41 |
| EW $\lambda$ = 0.90 | **0** | 4396.77 | 4396.77 |
| EW $\lambda$ = 0.925 | **0** | **4493.27** | 4493.27 |
| EW $\lambda$ = 0.95 | **0** | 4492.89 | 4492.89 |
| EW $\lambda$ = 0.975 | 2 | 4446.12 | 4491.03 |

### REFERENCES

[1] J. Duan, D. Shi, R. Diao, et al., "Deep-Reinforcement-Learning-Based Autonomous Voltage Control for Power Grid Operations," *IEEE trans. Power Syst.*, Early Access, 2019.

[2] R. Diao, Z. Wang, D. Shi, et al., "Autonomous Voltage Control for Grid Operation Using Deep Reinforcement Learning," *IEEE PES General Meeting*, Atlanta, GA, USA, 2019.

[3] H. Glavitsch, "Switching as means of control in the power system," *International Journal of Electrical Power & Energy Systems*, vol. 7, no. 2, pp. 92-100, 1985.

[4] A. A. Mazi, B. F. Wollenberg, M. H. Hesse, "Corrective control of power system flows by line and bus-bar switching," *IEEE trans. Power Syst.*, vol. 1, no. 3, pp. 258-264, 1986.

[5] E. B. Fisher, R. P. O'Neill, M. C. Ferris, "Optimal transmission switching," *IEEE trans. Power Syst.*, vol. 23, no. 3, pp. 1346-1355, 2008.

[6] A. Khodaei, and M. Shahidehpour, "Transmission switching in security-constrained unit commitment," *IEEE trans. Power Syst.*, vol. 25, no. 4, pp. 1937-1945, 2010.

[7] J. D. Fuller, R. Ramasra, and A. Cha, "Fast heuristics for transmission-line switching," *IEEE Trans. Power Syst.*, vol. 27, no. 3, pp. 1377-1386, 2012.

[8] P. Dehghanian, Y. Wang, G. Gurrala, et al., "Flexible implementation of power system corrective topology control," *Electric Power Syst. Research*, vol. 128, pp. 79-89, 2015.

[9] M. Alhazmi, P. Dehghanian, S. Wang, et al., "Power grid optimal topology control considering correlations of system uncertainties," *IEEE Tran. Ind Appl.*, Early Access, 2019.

[10] RTE France, ChaLearn, L2RPN Challenge. [Online]. Available: https://l2rpn.chalearn.org/

[11] D. Shi, T. Lan, J. Duan, et al., "Learning to Run a Power Network through AI," slides presentated at the 2019 PSERC Summer Workshop. [Online]. Available: https://geirina.net/assets/pdf/2019-PSERC_L2RPN%20Presentation.pdf

[12] R. S. Sutton, A. G. Barto, Introduction to reinforcement learning. MIT press Cambridge, vol. 2, no. 4, 1998.

[13] M. Lerousseau, A power network simulator with a Reinforcement Learning-focused usage. [Online]. Available: https://github.com/MarvinLer/pypownet

[14] V. Mnih, K. Kavukcuoglu, D. Silver, et al., "Playing atari with deep reinforcement learning," arXiv preprint arXiv:1312.5602, 2013.

[15] H. Van Hasselt, A. Guez, and D. Silver, "Deep reinforcement learning with double q-learning," in *30th AAAI Conference on Artificial Intelligence*, 2016.

[16] Z. Wang, T. Schaul, M. Hessel, et al., "Dueling network architectures for deep reinforcement learning," arXiv preprint arXiv:1511.06581, 2015.

[17] T. Schaul, J. Quan, I. Antonoglou, et al., "Prioritized experience replay," arXiv preprint arXiv:1511.05952, 2015.

[18] GEIRINA, CodaLab L2RPN: Learning to Run a Power Network. [Online]. Available: https://github.com/shidi1985/L2RPN.